\documentclass[a4paper]{revtex4}
\usepackage{graphicx}
\usepackage{fancyhdr}
\usepackage{amsmath}
\usepackage{color}
\usepackage{caption}

\usepackage{amsfonts}
\usepackage{amsmath,amssymb,amsthm}
\usepackage{mathrsfs}
\usepackage{float}
\usepackage{booktabs}
\usepackage{mathtools}

\usepackage[utf8]{inputenc}
\usepackage{caption,xspace}
\usepackage{multirow}
\usepackage[colorlinks,citecolor=blue]{hyperref}

\usepackage{verbatim}
\usepackage{float}
\usepackage{tikz}
\usetikzlibrary{shapes,arrows}

\begin{document}

\title{SUSY Implications from WIMP Annihilation \\into Scalars at the Galactic Center}


\author{Anibal D. Medina}
\affiliation{ARC Centre of Excellence for Particle Physics at the Terascale,\\
School of Physics, The University of Melbourne, Victoria 3010, Australia}

\begin{abstract}
An excess in $\gamma$-rays emanating from the galactic centre has recently been observed in the Fermi-LAT data. We investigate the new exciting possibility of fitting the signal spectrum by dark matter annihilating dominantly to a Higgs-pseudoscalar pair. We show that the fit to the $\gamma$-ray excess for the Higgs-pseudoscalar channel can be just as good as for annihilation into bottom-quark pairs. This channel arises naturally in a full model such as the next-to-minimal supersymmetric Standard Model (NMSSM) and we find regions where dark matter relic density, the $\gamma$-ray signal and other experimental constraints, can all be satisfied simultaneously.
Annihilation into scalar pairs allows for the possibility of detecting the Higgs or pseudoscalar decay into two photons, providing a smoking-gun signal of the model.

\end{abstract}

\maketitle

\thispagestyle{fancy}


%%%%%%%%%%%%%%%%%%%%%%%%%%%%%%%%%%
\section{Introduction}
Cosmic-ray experiments are a promising way to search for dark matter (DM). In particular, the satellite-based experiment Fermi-LAT is able to measure the $\gamma$-ray sky with unprecedented precision. If DM annihilates into photons with energies from $20$ MeV to $300$ GeV, an imprint can be left on these measurements. Intriguingly, an excess from the galactic centre (GCE) consistent with the range of DM density profiles indicated by observations and simulations of structure formation, has been identified in the Fermi-LAT data~\cite{Goodenough:2009gk,Daylan:2014rsa,Calore:2014xka}.

Taking the estimated uncertainty in the high-energy tail of the spectrum into account, DM annihilating to Higgs pairs close to threshold provides a good fit to the GCE~\cite{Calore:2014nla, Agrawal:2014oha}. In this work we analysed the new possibility of fitting the signal via DM annihilation into a Higgs and a pseudoscalar and study well-motivated DM models in which these channels arise naturally. We find that an acceptable fit is obtained for pseudoscalar masses up to around 150 GeV. In particular, in the region where the pseudoscalar is lighter than the Higgs, the fit  
is improved compared to the annihilation channel into Higgs-pairs~\cite{Gherghetta:2015ysa}. For a sufficiently light pseudoscalar, the fit becomes even better than that for the $b \bar b$-channel.

Within the context of supersymmetric models, the light pseudoscalar required for the Higgs-pseudoscalar channel is difficult to obtain in the minimal supersymmetric standard model (MSSM) due to collider and flavour constraints~\cite{Gherghetta:2014xea,Carena:2014nza},~\cite{Eriksson:2008cx}. Given these problems in the MSSM, we consider the NMSSM, where a light pseudoscalar is easier to obtain. Furthermore, additional contributions to the Higgs quartic coupling in the NMSSM alleviate the need for large stop-sector soft masses to increase the Higgs mass, improving the naturalness of the model~\cite{Gherghetta:2012gb}. Both the Higgs and the pseudoscalar can decay into two photons and although we find that the $\gamma$-ray line from pseudoscalar decays is distinguishable from the continuum only in an optimal-case scenario, more sensitive $\gamma$-ray experiments in the future may be able to detect it. Searching for these lines in the $\gamma$-ray spectrum could provide a smoking-gun signal of the new channels.

\section{The $\gamma$-ray excess from the galactic center }

We next discuss the annihilation channels to a Higgs-pseudoscalar pair ($ha$)  in the context of 
a two-Higgs-doublet model of type II with an arbitrary number of additional singlets. This includes the NMSSM which we focus on later.
We fix the Higgs  mass to $125$ GeV and its couplings to SM values, as implemented in \texttt{PYTHIA 8.201}~\cite{Sjostrand:2014zea}. We denote the pseudoscalar as $a$ and set $\tan\beta=3$ to fix its couplings. Furthermore, we assume that neither $a$ nor $h$ can decay into other scalars. 

\begin{figure}[t]
\centerline{
\includegraphics[height=5.2cm]{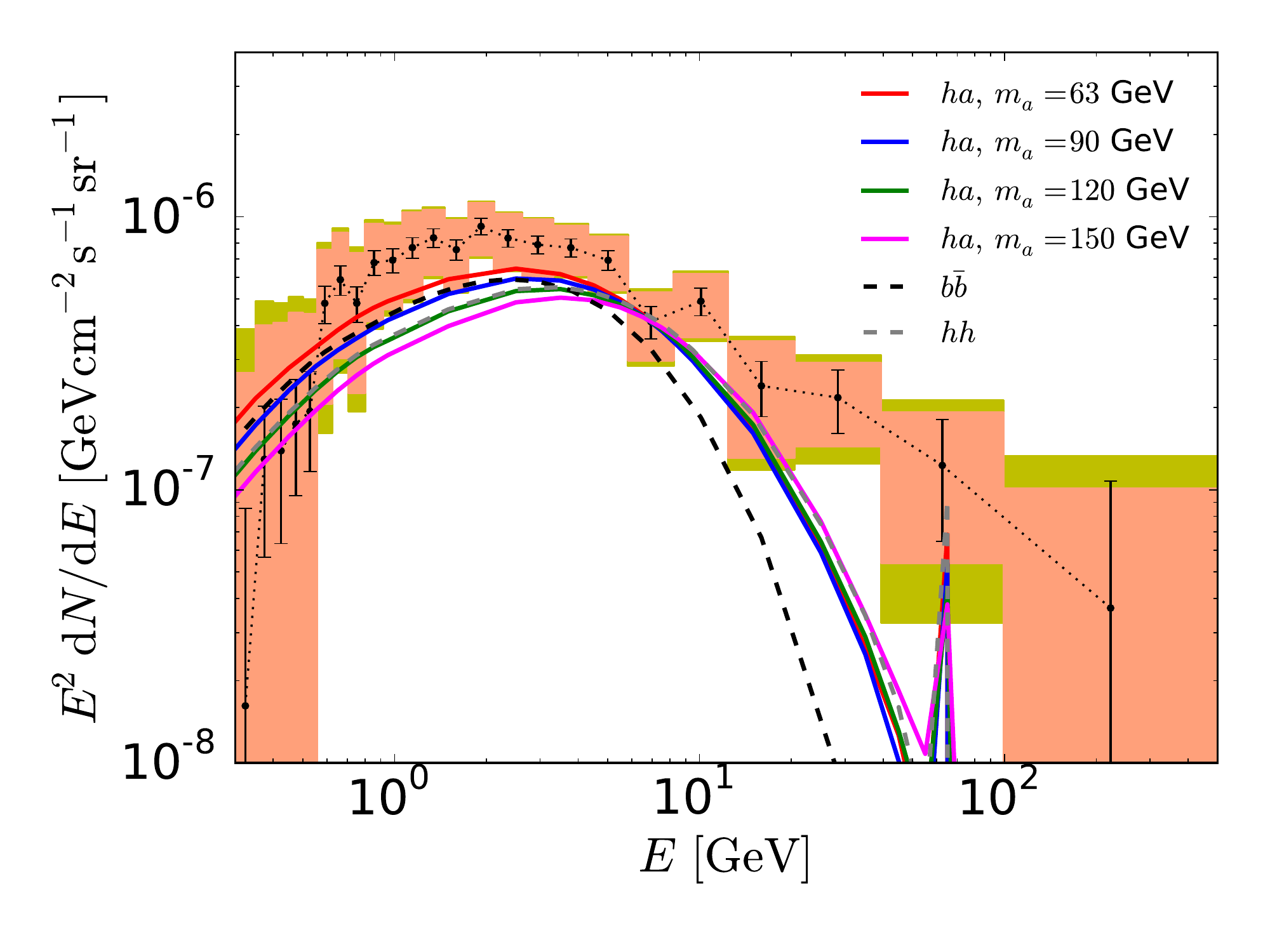}
\hspace*{3mm}
\includegraphics[height=5.2cm]{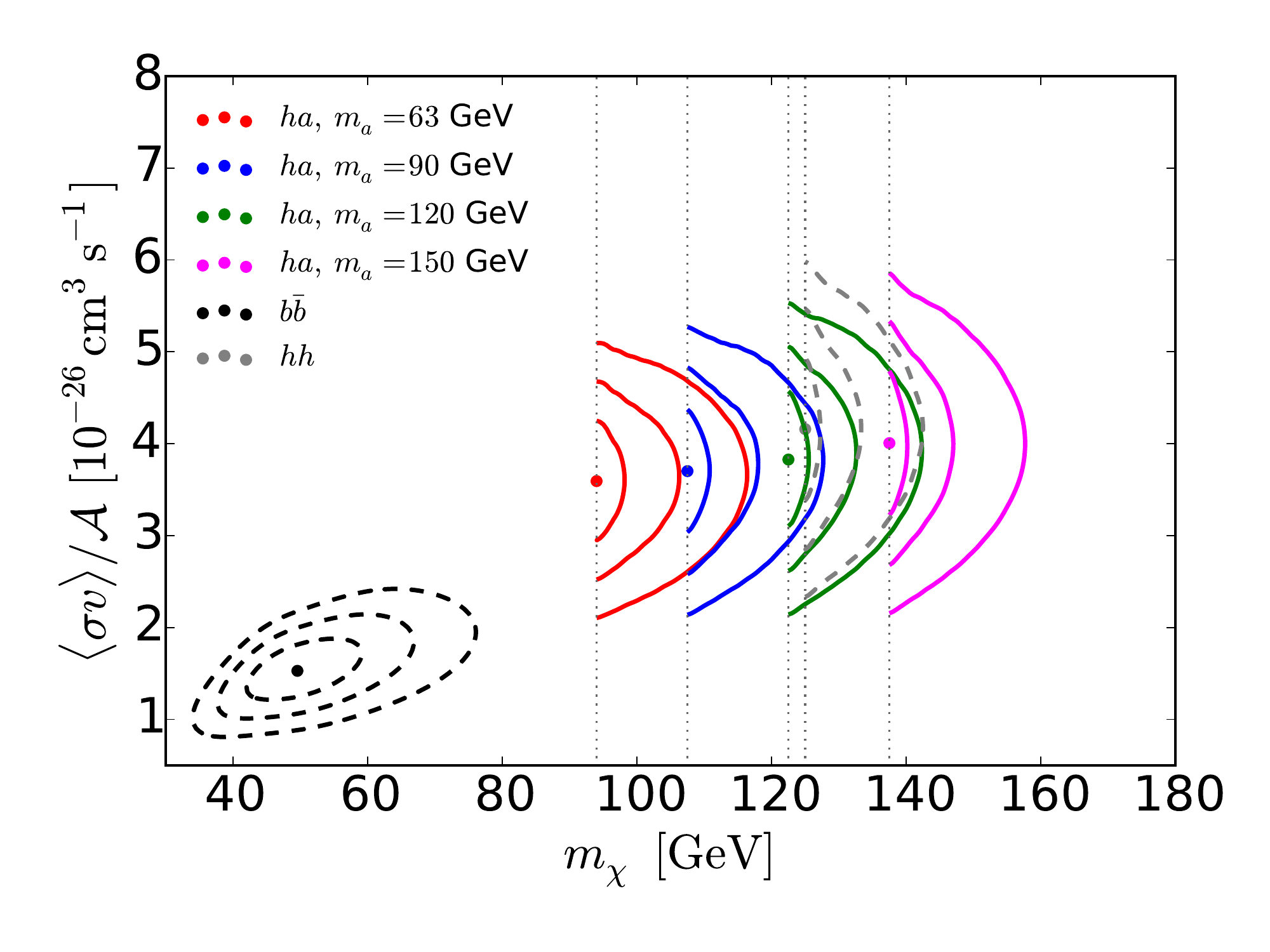}
}
\caption{(a) Spectrum of the GCE from~\cite{Calore:2014xka} and best-fit spectra for DM annihilation to $ha$ for different values of $m_a$. (b) Best-fit regions in the DM mass and cross section for $m_a=63, 90, 120$  and $150$ GeV, with contours delimiting the $1\sigma$-, $2\sigma$- and $3\sigma$-regions.}
\label{fig:fit-regions}
\end{figure}

We have performed our own fits to the reduced spectra of \cite{Calore:2014xka}. For completeness, we have also performed fits for the $b \bar b$- and $hh$-channels. We assume a generalised NFW profile,
\begin{equation}
\rho(r)=\rho_\odot \left(\frac{r}{r_\odot}\right)^{-\gamma}\left(\frac{1+r_\odot/R_s}{1+r/R_s}\right)^{3-\gamma}
\label{genNFW}
\end{equation}
with slope parameter $\gamma=1.26$, scale radius $R_s=20$ kpc and the DM density $\rho_\odot=0.4\, \mathrm{GeV}/\mathrm{cm}^3$ at the radial distance of the sun from the galactic centre $r_\odot$.
We use the prompt photon spectrum, $dN_\gamma/dE$, for annihilations into $b\bar b$ from \texttt{PPPC4MID}~\cite{Cirelli:2010xx} including electroweak corrections~\cite{Ciafaloni:2010ti} and find good agreement with our own simulation using \texttt{PYTHIA 8.201}~\cite{Sjostrand:2014zea}.  
We only consider the dominant decay channels of $a$ to $b\bar b$, $\tau^+\tau^-$, $c\bar c$, photons and gluons, 
and simulate the resulting prompt photon spectra for $hh$, $ha$ final states using \texttt{PYTHIA 8.201}~\cite{Sjostrand:2014zea}. 
Note that these spectra are unaffected by a possible singlet admixture of the Higgs or pseudoscalar. Indeed, such an admixture reduces their total decay widths, but the branching fractions remain unaffected (to leading order). The differential flux measured by Fermi-LAT is given by,
\begin{equation}\label{eq:dNdENFW}
\frac{dN}{dE}=\frac{\langle\sigma v\rangle_0}{8\pi \, m_{_{\rm DM}}^2} \frac{dN_\gamma^f}{dE} \int_{\rm l.o.s.} \hspace{-.2cm} ds\, \rho^2(r(s,\psi))
\end{equation}
with the line-of-sight integral $\int_{\rm l.o.s.}\hspace{-.1cm} ds$ over the squared DM density. The coordinate $r$ is centred on the galactic centre and can be expressed as $r^2(s,\psi)=r_\odot^2+s^2-2 r_\odot s \cos\psi$, where $s$ is the line-of-sight distance and $\psi$ is the aperture angle between the axis connecting the earth with the galactic centre and the line-of-sight. If DM annihilates into multiple final states, the different fluxes are summed over.

We use the reduced spectrum of the GCE from Ref.~\cite{Calore:2014xka} and the corresponding covariance matrix of the flux uncertainties including statistical and systematic errors which is publicly available. For simplicity we keep the pseudoscalar mass $m_a$ fixed and perform a two-parameter fit in the DM mass and annihilation cross section. In Figs.~\ref{fig:fit-regions} we show the resulting best-fit spectra from DM annihilation together with the spectrum of the GCE for different values of $m_a$ and the different annihilation channels. The smallest $m_a$ in these figures is chosen such that Higgs decays to pseudoscalars is kinematically forbidden, i.e. $m_a > m_h/2$. The salmon-colored boxes depict the empirical model systematics~\cite{Calore:2014xka}, the error bars correspond to the statistical errors, and the yellow boxes are the combination of the statistical errors, empirical model systematics and other systematics. 

Notice that, similar to the $hh$-channel, the spectra for the $ha$-channel in Fig.~\ref{fig:fit-regions} have a peak at energies $m_h/2$ which is produced from on-shell decays of the Higgs to two photons. The peak is less pronounced than for the $hh$-channel because there is only one Higgs in the final state. There is only mild line broadening for the best-fit masses, since the $hh$- or $ha$-pair is produced close to threshold. 
Notice also that the spectra have no visible peak at energies $m_a/2$ from pseudoscalar decays to two photons.  Indeed, the peak in Fig.~\ref{fig:fit-regions} for the $ha$-channel is purely due to Higgs decays. 
Using the Fermi-LAT limits on $\gamma$-ray lines~\cite{Ackermann:2013uma},  it was estimated that the line strength from the decay $hh\to 4\gamma$ is just below current limits and may be detected (or excluded) in the near future~\cite{Calore:2014xka}. In contrast, the intensity of the line produced by pseudoscalar decays is too weak to be in tension with line searches. We find that only in the best-case scenario of $m_a\sim 150$ GeV and $\tan\beta\sim 1$ is the peak barely distinguishable from the continuum. More sensitive $\gamma$-ray experiments may be able to detect the photon peak for lower $m_a$ and larger $\tan\beta$ in the future.

In Fig.~\ref{fig:fit-regions} we also show the best-fit regions in the DM mass and cross section for the $ha$-channel. 
For comparison, we also show the best-fit regions that we find for the $b \bar b$- and $hh$-channels. 
Similar to~\cite{Calore:2014nla, Agrawal:2014oha}, we assume a multiplicative astrophysical-uncertainty factor $\mathcal{A}$ for our best-fit cross-sections.
This factor takes into account the uncertainties in the local DM density $\rho_\odot$, the scale radius $R_s$ and the slope parameter $\gamma$. 
We have used the same reference values for $\rho_\odot$, $R_s$ and $\gamma$ for our fits as~\cite{Calore:2014nla} and we therefore use their estimate for the range of the astrophysical-uncertainty factor, $\mathcal{A} \in [0.17,5.3]$.
Notice from Fig.~\ref{fig:fit-regions} that the best-fit regions for the $ha$-channel 
lie very close to threshold. The best-fit cross sections are fairly independent of the pseudoscalar and DM mass.  
In Table~\ref{tab:bestfit}, we show the values and 1$\sigma$-regions for the best-fit DM mass and cross section and the associated $\chi^2$ and $p$-values for the different annihilation channels. The smaller $m_a$ lead to a better fit, i.e.~a smaller $\chi^2$ and thus a larger $p$-value. In particular, for $m_a\lesssim 120$ GeV, the fit for the $ha$-channel is better than for $hh$ final states and for $m_a=63$ GeV the fit becomes better than for $b\bar{b}$ final states. It can also be seen that the best-fit cross sections for the scalar channels are larger than for the $b \bar b$-channel.

%
%\begin{figure}[tbp]\centering
%\begin{subfigure}{0.49\linewidth}
%\includegraphics[width=\linewidth]{fig1a}
%\caption{\label{fig:fit-spectrum}} 
%\end{subfigure}
%\hfill
%\begin{subfigure}{0.49\linewidth}
%\includegraphics[width=\linewidth]{fig1b}
%\caption{\label{fig:fit-regions}} 
%\end{subfigure}
%\caption{(a) Spectrum of the GCE from Ref.~\cite{Calore:2014xka} and best-fit spectra for DM annihilation to $ha$ for different values of $m_a$. (b) Best-fit regions in the DM mass and cross section for $m_a=63, 90, 120$  and $150$ GeV, with contours delimiting the $1\sigma$-, $2\sigma$- and $3\sigma$-regions.}
%\end{figure}

\begin{table}[btp]\centering
\begin{tabular}{lccccc}
\hline
channel& $m_a$ [GeV] & $m_{_{\rm DM}}$ [GeV] & $\langle\sigma v\rangle_0 \, [10^{-26}\mathrm{cm^3}/\mathrm{s}]$ & $\chi^2_{\mathrm{min}}$ & $p$-value\\
\hline
$b\bar b$ & & $49.6_{-6.3}^{+8.1}$ & $1.5_{-0.2}^{+0.3}$ &  $24.5$ & $0.32$ \\
$hh$ & & $125.0_{-0.0}^{+2.3}$ & $4.2_{-0.8}^{+0.8}$ &  $30.0$ & $0.12$ \\
\hline
\multirow{4}{*}{$ha$}
 &$ 63$ & $94.0_{-0.0}^{+4.2}$ & $3.6_{-0.6}^{+0.7}$ &  $22.4$ & $0.43$ \\
 &$ 90$ & $107.5_{-0.0}^{+3.4}$ & $3.7_{-0.7}^{+0.7}$ &  $25.3$ & $0.28$ \\
 &$120$ & $122.5_{-0.0}^{+3.0}$ & $3.8_{-0.7}^{+0.8}$ &  $30.3$ & $0.11$ \\
 &$150$ & $137.5_{-0.0}^{+2.7}$ & $4.0_{-0.8}^{+0.8}$ &  $36.0$ & $0.03$ \\
\hline
\end{tabular}
\caption{
Best-fit values and $1\sigma$-regions from our fits for the $b \bar b$, $hh$, $ha$  annihilation channels.}
\label{tab:bestfit}
\end{table}

Let us briefly mention that there is no dedicated analysis for annihilation into $hh$ and $ha$ in the preliminary dwarf-spheroidal galaxy searches. However, $b$-quarks are still the dominant decay product for these channels.  We obtain the rough estimate $\langle \sigma  v \rangle_0  \lesssim 3\times 10^{-26}\text{cm}^3/\text{s}$ for DM with $m_{_{\rm DM}} \sim 100$ GeV annihilating into $hh$ or $ha$ close to threshold, depending relatively weakly on $m_{a}$ in the range of interest here.
%We can therefore use the limits for the $b \bar b$-channel by observing that DM of mass $m_{_{\rm DM}}$ annihilating into four $ b$-quarks behaves kinematically similar to DM of mass $m_{_{\rm DM}}/2$ annihilating into two $ b$-quarks. For heavier DM masses the bound on $\langle \sigma  v \rangle_0$ is weaker.

%The DM interpretation of the GCE is also constrained by other astrophysical observations, in particular by measurements of the anti-proton flux and radio signals from the galactic centre. However, it has been argued that systematic uncertainties in the modelling of both the expected anti-proton flux \cite{Cirelli:2014lwa} and the radio signals \cite{Cholis:2014fja} were underestimated in Ref.~\cite{Bringmann:2014lpa} which focuses on the required annihilation cross section for the $c \bar c$- and $b \bar b$-channel. Taking these uncertainties into account, the required cross sections may still be consistent. Once again, no dedicated analysis into $hh$ or $ha$-channel has been performed so far.

\section{Annihilation into $ha$ in the NMSSM}

\begin{figure}[t]
\centerline{
\includegraphics[height=2.2cm]{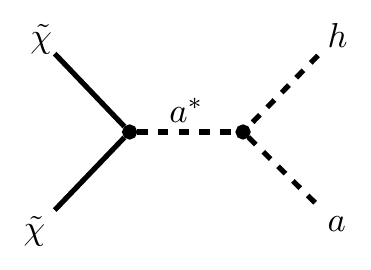}
\hspace*{3mm}
\includegraphics[height=2.2cm]{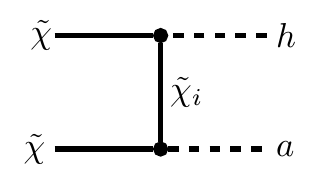}
}
\caption{Feynman diagrams contributing to neutralino annihilation to $ha$-pairs. Note that there is a $u$-channel diagram in addition to the $t$-channel one.}
\label{fig:XXhadiag}
\end{figure}

\begin{figure}[t]
\centerline{
\includegraphics[height=5.5cm]{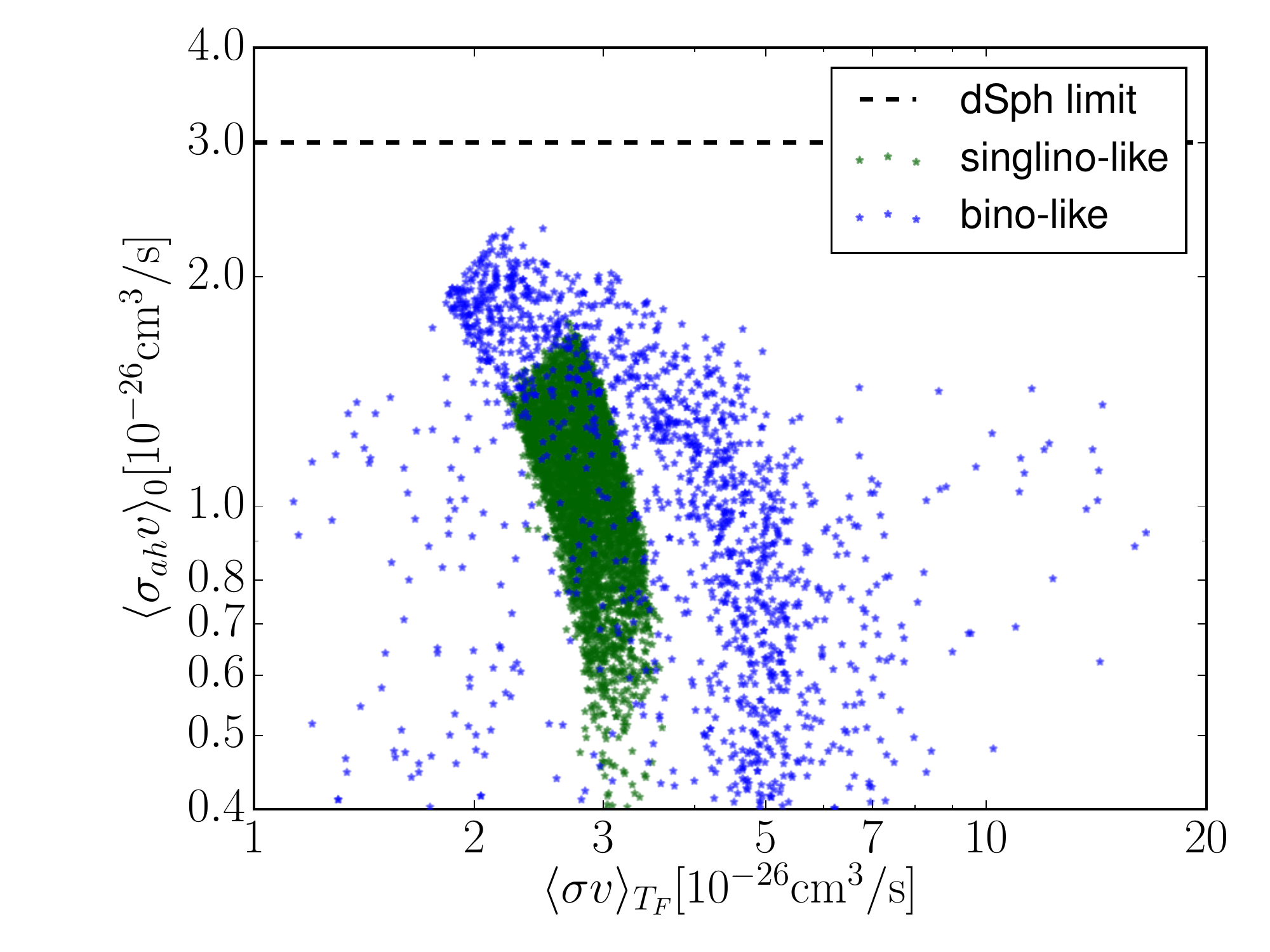}
\hspace*{3mm}
\includegraphics[height=5.5cm]{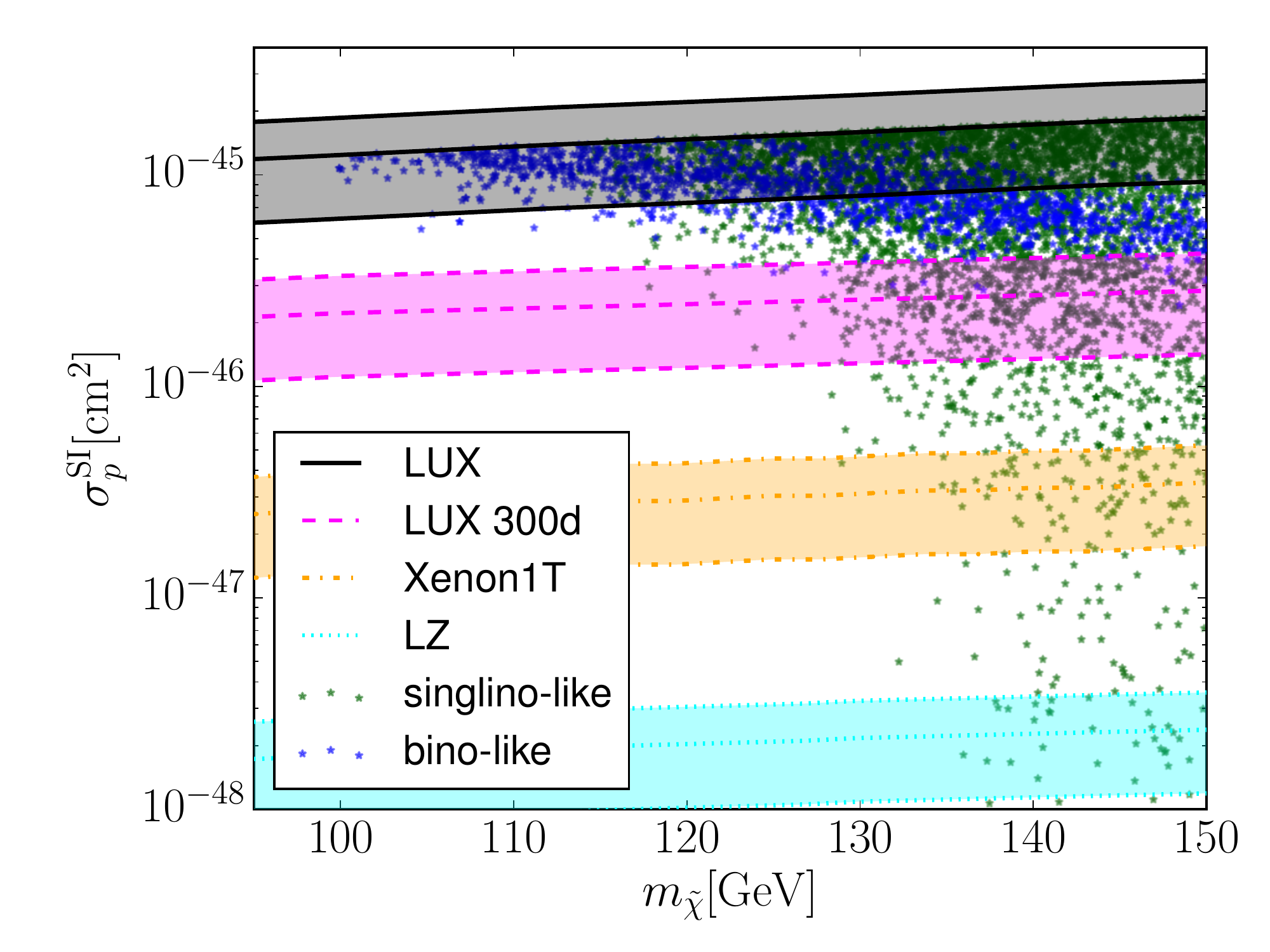}
}
\caption{The left panel (a) shows the annihilation cross section at the present time, $\langle \sigma_{ha} v\rangle_0$, and at freeze-out, $\langle\sigma v \rangle_{T_F}$. The right panel (b) shows the spin-independent cross section for scattering of LSPs off nucleons versus the LSP mass.}
\label{fig:HAscanLSP2}
\end{figure}

The annihilation of neutralinos is $p$-wave suppressed if the final state is even under $CP$, such as for $hh$ or $aa$~\cite{Griest:1989zh}. Therefore it is only through the $ha$-channel that neutralino annihilation into scalars can account for the GCE. The graphs that contribute to the corresponding cross section are the $s$-channel exchange of the two pseudoscalars and the $t/u$-channel exchange of the neutralinos. These are shown in Fig.~\ref{fig:XXhadiag}. We have performed two random scans using \texttt{NMSSMTools\;4.4.0}~\cite{Ellwanger:2004xm} for the $ha$-channel, one optimised for bino-like LSPs and the other for singlino-like LSPs. The masses of the squarks, sleptons, gluino and wino are fixed at 2 TeV and the remaining free parameters are scanned over the ranges shown in Table~\ref{tab:HAscan}. Collider constraints and direct and indirect detection constraints are all satisfied. Here $A_\lambda$ is partly determined by the requirement that the singlet admixture to the Higgs is small. In Fig.~\ref{fig:HAscanLSP2}, we show a scatter plot of the annihilation cross-sections during freeze-out and at late times. 
The approximate degeneracy of these cross sections  reflects the dominance of the $\tilde{\chi}\tilde{\chi}\rightarrow ha$ process and the absence of a large special enhancement such as a resonance. In Fig.~\ref{fig:HAscanLSP2}, we also show a scatter plot of the cross section for spin-independent DM-nucleon scattering versus the DM mass. The LUX collaboration will test a considerable portion of the singlino-like points and almost the entirety of the bino-like points from our scan of the $ha$-channel. The vast majority of points with a singlino-like LSP would then be probed by XENON1T, although we find a few such points that would evade even the projected LZ limits. 

\begin{table}[bt]\centering
\begin{tabular}{ l  c  c  c ccccc}
\hline
& $\Delta A_\lambda$ [GeV] & $A_\kappa$  [GeV] & $\mu_{\rm eff}$  [GeV] &  $M_1$  [GeV] &$\lambda$ & $\kappa$ & $\tan \beta$ \\
\hline
bino-like & [-50,50] & [-100,100] & [-300,-100]  & [60,170] & [0.6,1.4] & [0.1,1.6] & [2,5]  \\
singlino-like &  [-50,50] & [-100,100] & [-600,-200] & 2000 & [0.6,1.4] & [0.05,0.5] & [2,5] \\
\hline
\end{tabular}
\caption{Parameter ranges for the two random scans for the $ha$-channel. The first and second line are for the scan optimised for bino-like and singlino-like LSPs, respectively.
}
\label{tab:HAscan}
\end{table}

%\begin{figure}[bt]\centering
%\begin{subfigure}{0.49\linewidth}
%\includegraphics[width=\linewidth]{fig4a}
%\caption{\label{fig:HAscanLSP1}}
%\end{subfigure}
%\hfill
%\begin{subfigure}{0.49\linewidth}
%\includegraphics[width=\linewidth]{fig4b}
%\caption{\label{fig:HAscanLSP2}}
%\end{subfigure}
%\caption{The left panel (a) shows the annihilation cross section at the present time, $\langle \sigma_{ha} v\rangle_0$, and at freeze-out, $\langle\sigma v \rangle_{T_F}$. The right panel (b) shows the spin-independent cross section for scattering of LSPs off nucleons versus the LSP mass.}
%\label{fig:HAscanLSP}
%\end{figure}

\section{Conclusions}
We have investigated the possibility of explaining the $\gamma$-ray excess at
the center of the galaxy via DM annihilation into a Higgs and a pseudoscalar, and 
showed that a good fit of the photon spectrum can be accomplished. We have also
shown that this possibility arises naturally in the context of the NMSSM where
regions exist that satisfy all experimental constraints.

\begin{acknowledgments}
The author would like to thank to all of organizers of HPNP2015 and great hospitality during the conference. This work was supported in part by the Australian Research Council.

\end{acknowledgments}

\bigskip % extra skip inserted

\end{document}